\RequirePackage{fix-cm}
\documentclass[twocolumn]{svjour3}          
\smartqed  
\usepackage{graphicx}
\usepackage{mathptmx}      
\usepackage{mathrsfs,amsmath}
 \journalname{Preprint}
\begin{document}

\title{Taming chaos with active grids
}
\subtitle{Reproducibility in turbulent wind tunnel experiments}


\author{Lars Kr\"oger         \and
	    Michael H\"olling     \and
        Gerd G\"ulker         \and
        Joachim Peinke
}


\institute{ForWind and Institute of Physics, University of Oldenburg \at
              K\"upkersweg 70, 26129 Oldenburg, Germany \\
              Tel.: +49-441-798-3514\\
              Fax: +49-441-798-5099\\
              \email{lars.kroeger@uni-oldenburg.de}           
}

\date{Received: date / Accepted: date}

\maketitle

\begin{abstract}
A wind tunnel experiment is shown to examine the reproducibility of generated wind fields in the laboratory  utilizing an active grid. A motion protocol of the grid, designed to mimic a measured atmospheric wind speed time series, is therefore repeated 50 times and the flow is measured by a 2D hotwire (x-wire) system. The acquired wind speed time series are analysed in two different ways to identify the reproducible time scales. A direct comparison using the cross covariance function combined with two filtering methods and a statistic approach using one and two point analysis. Like that the reproducible time and length scales could be linked to the active grid motion and design. The identification of the reproducible time scales of a turbulent signal generated by means of an active grid is suitable to improve the possible wind fields in laboratory experiments.
\end{abstract}

\section{Introduction}
\label{intro}
The omnipresence of turbulence in nature and various processes like mixing and fluid structure interactions are increasing the demand on experiments under controlled and pre-defined turbulent conditions. This is especially true for research in the field of wind energy, where experiments in situ are always performed in a mixture of different and constantly changing atmospheric phenomena. Thus, experiments under controllable laboratory conditions allow to reduce the complexity of the acting inflow what is of major importance of many problems. 
Commonly turbulent fields in a wind tunnel experiment are mostly realized by the use of passive elements like grids and stakes \cite{Roach1986},\cite{Batchelor1993}. The simplicity of passive elements is satisfying the demand of a decreased complexity of the acting inflow but is also limiting the experiments to one characteristic flow field for every combination of available elements in the laboratory. To enhance the possibilities of the generation of turbubulence in the wind tunnel the so called active grid, first introduced by Makita \cite{Makita1991}, are used. Dynamically driven parts like shafts with flaps  generate a variety of turbulent wind fields without a change of the experimental setup in the wind tunnel \cite{Mydlarski2017}. The so far used active motion is often fully random and generating statistically well defined, homogeneous and isotropic turbulence \cite{MydlarskiWarhaft1996}, \cite{Larssen2011}. \\
Having wind energy research in mind the utilization of active grids is an excellent method to generate user defined wind fields with lots of benefits. The turbulent conditions of interest interacting with a wind energy converter (WEC), namely the atmospheric boundary layer (ABL), are very characteristic and focus of a lot of research \cite{waechter2012}. The increased degree of freedom allows active grids to generate flows with several of these special characteristics for example shear flows \cite{Cekli2010} \cite{Hearst2017}, high turbulence intensities, intermittency \cite{Knebel2011} or high Taylor Reynolds numbers \cite{Neuhaus2020}.\\
 For wind energy research experiments should depict the turbulent and gusty nature of flow fields in the ABL as accurate as possible. Furthermore and no less important the generation of specific flow fields should be reproducible since many experiments have to be executed several times with the same flow fields. The possibility to rerun the motion protocol of the active grid shafts can be exploited to achieve that reproducibility in wind tunnel experiments. Moreover flow fields can be generated presenting a special flow situation like a gust where the temporal appearance is exactly predefined or scaled to the length scales of the respective test object. This provides perfect conditions to thoroughly investigate aerodynamic behaviour or validate benefits of passive and active control mechanisms of WECs.\\
In this study the reproducibility of active grid generated wind fields is shown. An experiment is presented facilitating an active grid to generate an exemplarily time-wise downscaled wind speed time series with atmospheric-like features. This time series is afterwards generated several times by repeating the active grid motion protocol and measured by a hotwire system. The resulting wind speed time series are then analysed in detail focusing on two aspects: the reproduction of direct structures of the signal in time (imprinted signal) and the reproducibility of statistical features of the turbulence (background turbulence).\\
This paper is divided in three different sections. In the next section the experimental setup, the process to design an active grid motion protocol and the atmospheric wind data set are described. The atmopheric wind data is used as the base to downscale the atmospheric-like features. Section \ref{sec_results} contains the discussion of the measured data to determine the reproducibility of generated turbulent wind speed time series, including a brief comparison to determine the possibility to downscale the atmospheric wind speed time series in a wind tunnel experiment. Finally the results are summarized and a conclusion is given.
\section{Experimental setup}
\label{sec:1}
In this section a detailed description of the experimental setup is given. It starts with the used wind tunnel and active grid system to generate the investigated turbulent wind field, followed by a short overview of the hotwire hardware used to measure the wind speed. Finally the design process and concept of a motion protocol to recreate an atmospheric wind speed time series in the wind tunnel is explained.
\subsection{Wind tunnel and active grid}
\label{sec:2}
\begin{figure}
  \includegraphics[width=.49\textwidth]{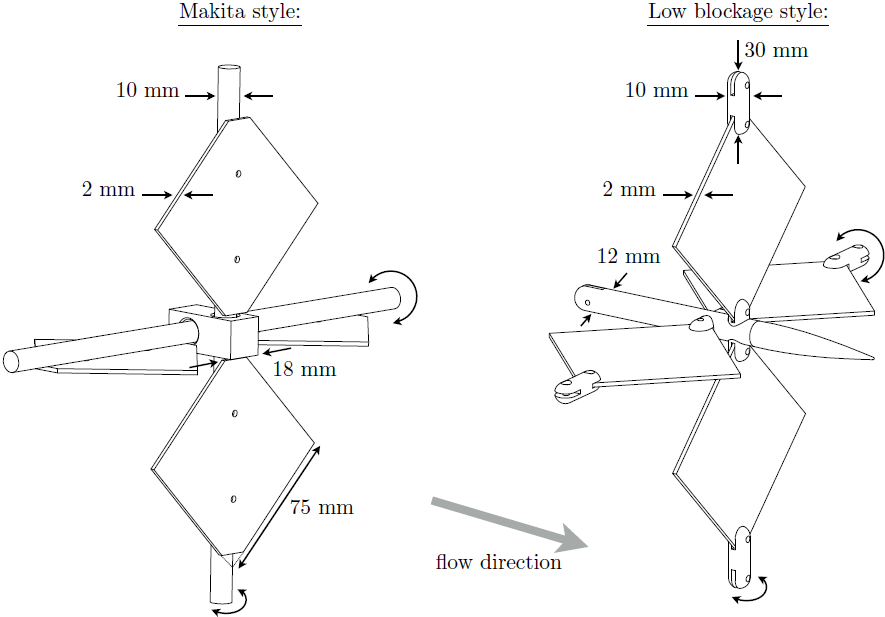}
\caption{Active grid design: Makita style vs. low blockage style. \cite{Weitemeyer2013}}
\label{pic:design}       
\end{figure}

The experiment was carried out in a wind tunnel of the University of Oldenburg. It is a wind tunnel of the 'G\"ottingen'-design type and has a cross section of $0.8 \times 1~m^{2}$ (height x width). It can be operated at wind speeds between $(1-50)~\frac{m}{s}$. Experiments can be done both in open or closed test section configuration. In this study the open configuration was used. The turbulence intensity of this configuration without a further set-up is below $0.3~\%$\\
\begin{figure}
  \includegraphics[width=.49\textwidth]{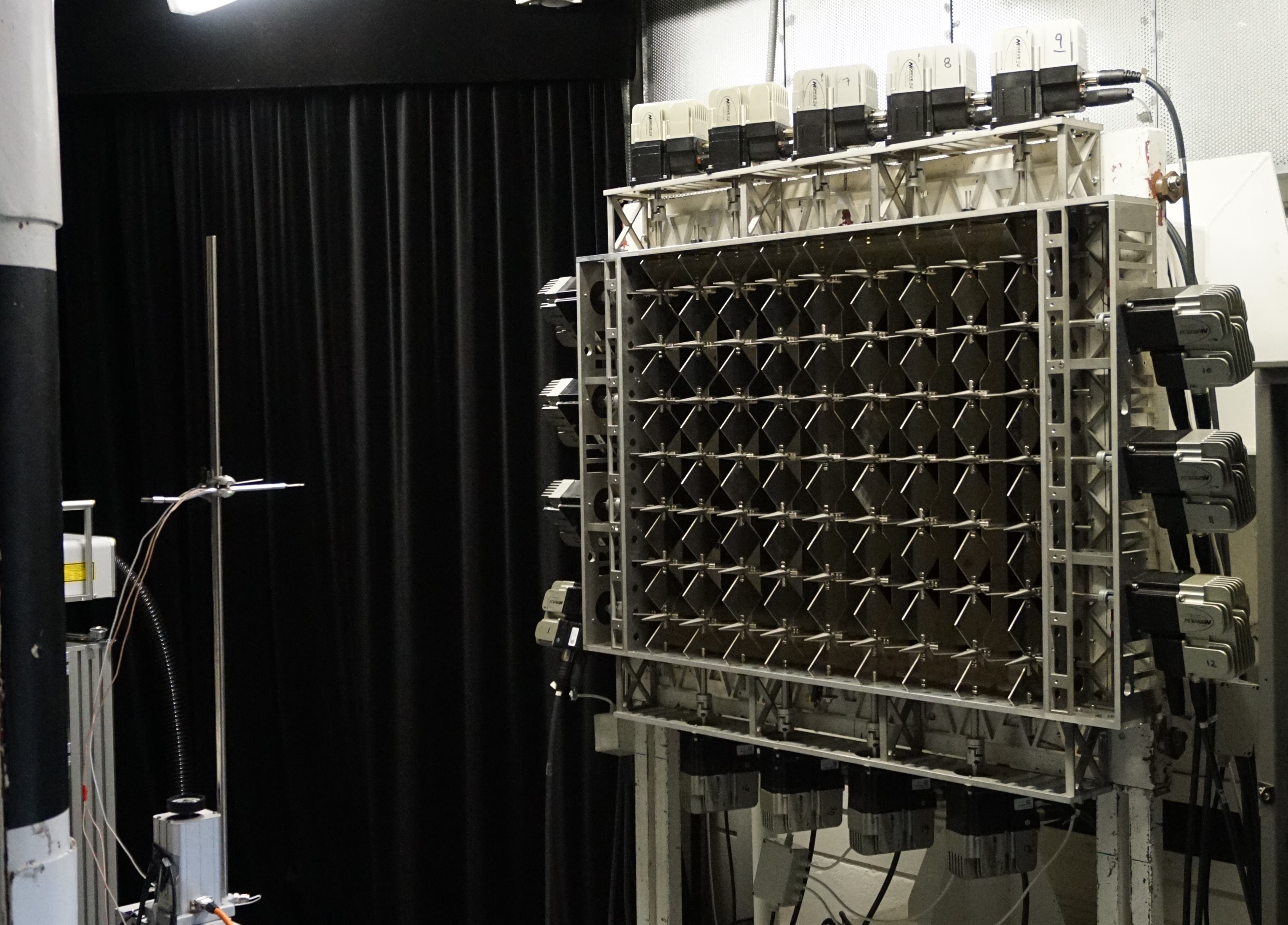}
\caption{Active grid mounted to the wind tunnel outlet. In ten mesh width distance a 2D hotwire probe is placed to measure the generated active grid wake.}
\label{pic:grid}       
\end{figure}
A picture of the set-up can be seen in figure \ref{pic:grid},where a part of the open test section is shown. On the right the active grid mounted to the wind tunnel outlet can be seen and on the left a 2D hotwire probe. The active grid used is composed of 16 shafts, nine in vertical and seven in horizontal orientation, thus resulting in a mesh width M of $11~cm$. Stepping motors are attached to every shaft leaving them to be individually controllable via a National Instruments motion control system.\\
 One major difference to the first active grid introduced by Makita is a low blockage design shown in figure \ref{pic:design} \cite{Weitemeyer2013}.
Both design styles are presented exemplarily for an intersection of the moving shafts. On the left the old Makita style is shown. It is consisting of continuous rods which are hold in place at the junction by a square block. The flaps are fixed on top of the rods by screws. For the new low blockage design on the right the square flaps are connected by short pieces of rod and the intersections of the shafts are mantled by a streamlined casing. This way the minimal blockage of the cross section when the grid is fully opened could be decreased from around $25~\%$ to values below $5~\% $, while the maximum blockage around $90~\% $ for both design types stays the same. Note that this low blockage design is important to enhance the possibility to generate predefined flow structures by the grid. A grid with a higher level of blockage generates also a higher level of background turbulence \cite{Roach1986} which is superposing the imprinted signal, which limits the possibility to modulate the flow by special motions of the flaps in the corresponding frequency range. A reduction of the background turbulence induced by the active grid design improves the variety of wind fields which can be generated. This reduction is a topic of ongoing research in our group but will not be part of this study.

\subsection{Measurement System}
A Dantec Dynamic 2D hotwire system was used to measure the generated flow fields. The hotwire probe was located 10M ($1.1~m$) downstream at the centerline of the active grid. The sampling frequency of the AD-converter was set to 20 kHz and the analog signal was low pass filtered at 10 kHz to avoid aliasing. The setup can be seen in figure \ref{pic:grid}. Synchronization of the active grid motion with the data acquisition is realized using a trigger signal send out by the motion control system of the grid. 

\subsection{Motion protocol} 
\begin{figure}
  \includegraphics[width=.4\textwidth]{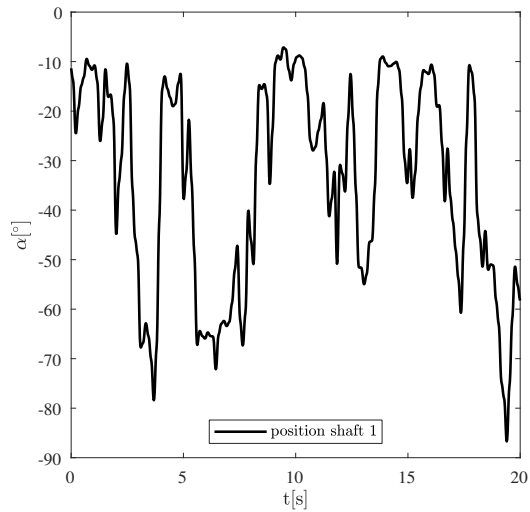}
\caption{Extraction of a motion protocol. Shown is the absolut flap angle position of one shaft.}
\label{pic:excitation}       
\end{figure}

\begin{figure}
  \includegraphics[width=.4\textwidth]{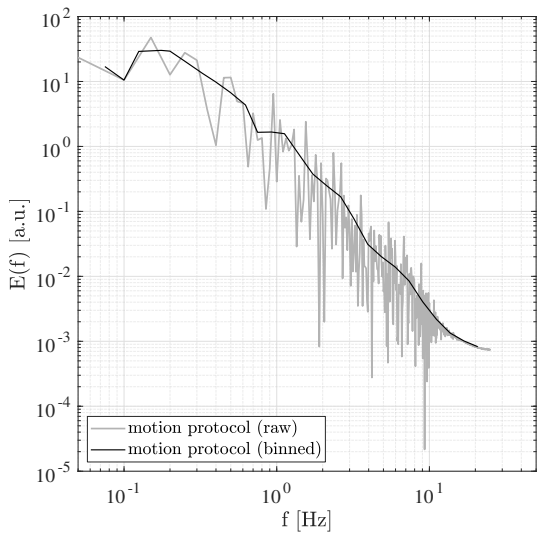}
\caption{The power spectrum of the flap angle time series of one active grid shaft.}
\label{pic:spec_excitation}       
\end{figure}

The properties of the generated turbulence downstream of the active grid is influenced mainly by two factors. The design of the grid (e.g. mesh width, minimal blockage) is defining a background turbulence. The change of the angle of the flaps with respect to the inflowing wind is resulting in a change of the blockage and redirection of the flow. This can be used to increase the turbulence in the active grid wake. A specific variation of the flap angle $\alpha$ over time is called in the following a motion protocol. Such a motion protocol is a matrix containing the  absolute flap angle $\alpha$ per time step for every one of the 16 shafts. Figure \ref{pic:excitation} shows an exemplarily extraction of such a time series for one of the shafts, which allows an individually defined movement of each of the shafts. As a further feature of our set up a TTL levelled trigger signal synchronized to the movements can be used to synchronize the measurements.\\
For wind energy research a main requirement of turbulent wind fields in wind tunnel experiments is the downscaling of atmospheric-like turbulence to the laboratory scale. The scaling factor hereby depends on the size and type of the interacting test object with the wind for example a model wind turbine. For the downscaling we make use of Taylor's hypthesis of frozen turbulence, thus we only consider the time. The wind speed magnitude is not changed. A downscaling of the fluctuations of the wind speed time series to match a model wind turbine is realized by speeding up the time series in the wind tunnel. As an example for the downscaling of the model wind turbine Oldenburg (MoWiTO 1.8) \cite{Berger2018} to the NREL 5MW reference turbine we obtain a linear scaling factor of about 68.\\
The main challenge now is to create such a motion protocol which produces a predetermined wind speed time series in a wind tunnel. In the following we present a method introduced by Reinke \cite{reinkediss} to reproduce a specific wind speed time series measured in the atmospheric boundary layer by a LiDAR system. Having determined the scaling factor the original time series is translated into a motion protocol using a so called transfer function. The transfer function is obtained by measurements of the relation between different flap angles and the observed wind signal at a position of interest in the wind tunnel. By measuring short time intervals for all different angles with the hotwire system we obtain a look-up-table containing the mean wind speeds $\langle u \rangle$ and the standard deviations $ \sigma_{u}$ for the respective combination of flap angles. The look up table allows to construct a motion protocol to generate a predefined time series using the superposition principle. A further and more detailed discussion of the transfer function and the generation of turbulent wind speed time series can be found in \cite{Neuhaus2021}.\\
In this study a set of LiDAR measurements is translated into a motion protocol with atmospheric-like fluctuations. The acquisition of the LiDAR data and its post processing are described in detail as part of the Smart Blades project report \cite{SmartBlades}. The absolut flap angle time series for one of the shafts is shown in figure \ref{pic:excitation}. Additionally the frequency spectrum of the flap angle time series is shown in figure \ref{pic:spec_excitation} to give an impression at which time scales energy is fed into the flow. The frequency spectrum is shown unfiltered and binned. The binning process for the frequency spectrum is explained in appendix \ref{sec_methods_analysis}. Energy is brought into the flow for this example at frequencies of $0.1~Hz$ to $25~ Hz$.\\
The wind tunnel speed was set to the mean wind speed of the LiDAR wind speed time series and the motion of the grid inducing the wind speed fluctuations will imprint the temporal evolution 68 times faster compared to the original. In the further investigation the time will be displayed in the laboratory time scale. Thus, the time of original LiDAR data has to be divided by a factor of 68.\\

\section{Results}\label{sec_results}
In this section the measured wind speed time series are analysed regarding the signal structure and the statistical behaviour. First the generated wind speed time series is compared briefly to the atmospheric original (\ref{sec_downscaled}), followed by the discussion of the signal structure over time of the repeatedly measured wind speed time series (\ref{sec_crosscovariance}). Finally the data is analysed by statistic means (\ref{sec_statistic}). The description of the analytic tools used for this discussion can be found in the appendix \ref{sec_methods_analysis}. 

\subsection{Downscaled wind speed time series} \label{sec_downscaled}
\begin{figure}
	\includegraphics[width=.4\textwidth]{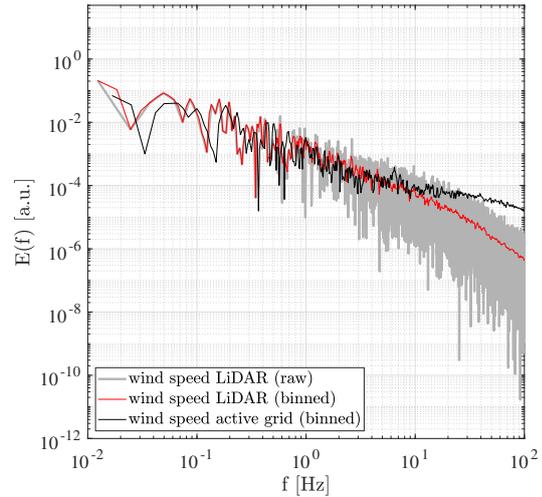}
	\caption{The power spectrum of the atmospheric wind speed time series and the recreated wind speed time series in the wind tunnel given in units of the laboratory frequency.}
	\label{pic:spec_LiDARvsgrid}       
\end{figure}

\begin{figure}
  \includegraphics[width=.4\textwidth]{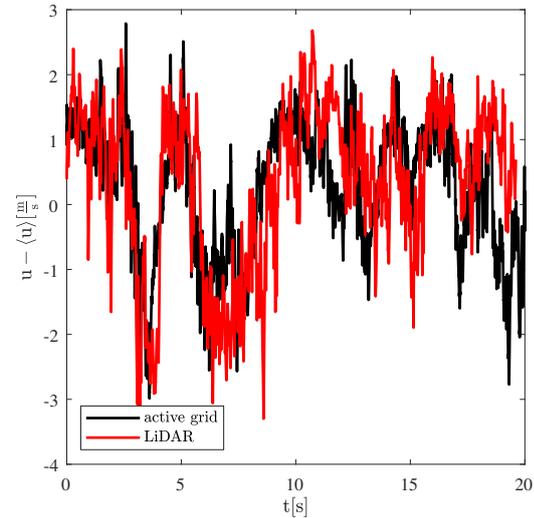}
\caption{The atmospheric wind speed time series and the generated wind speed time series in the wind tunnel given in the frame of laboratory time.}
\label{pic:ts_LiDARvsgrid}       
\end{figure}

In a first step, a brief comparison of the generated wind speed time series measured in the wind tunnel to the original should be given. The aim was to generate a wind speed timeseries with fluctuations comparable to wind speed data measured in the ABL and the downscaling in time. To check the reproduciblity not the perfect recreation of the data is necessary. Instead, we want to show that a very realistic wind field, which is useful for further investigations, can be generated quite easily.\\
With the active grid motion protocol frequencies between $0.1~Hz$ to $25~Hz$ are imprinted onto the flow (fig. \ref{pic:spec_excitation}). So a high resemblance is expected in this frequency range. In figure \ref{pic:spec_LiDARvsgrid} frequency spectra of both time series are shown in the range between $10^{-2}~Hz$ to $100~Hz$ in the laboratory scale and a good match for frequencies up to $25~Hz$ can be examined. A frequency of $25~Hz$ at a wind speed of $12~\frac{m}{s}$ results in a structure of about $50~cm$ imprinted by the active grid. Thus the wind dynamics on the scale of about $50~cm$, which is in the range of our used model wind turbine rotor, can be designed and repeated accurately.\\
For a direct comparison of the generated time series and the original data both time series are shown in figure \ref{pic:ts_LiDARvsgrid}. The fluctuations of the wind speed are plotted over the laboratory time in seconds measured by the LiDAR and the hotwire system. Both time series are filtered by a moving average serving as a low pass filter at $25~Hz$. 
 Filtered like this a maximal cross covariance coefficient of $0.53$ is reached when correlating the two time series. We can see that the basic trend can be recreated quite well as the dynamic features of the generated time series on the larger time scales are following the original data. 

\subsection{Determining reproducibility based on the cross covariance funktion}\label{sec_crosscovariance}
\begin{figure}[htb]
	\includegraphics[width=.49\textwidth]{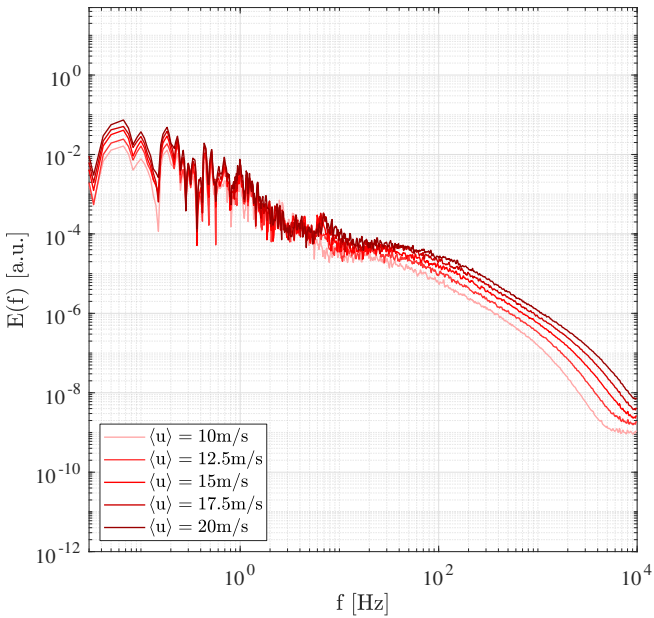}
	\caption{The turbulent power spectrum of the active grid generated wake for five different wind speeds smoothed by a windowing filtering.}
	\label{pic:spectrum_rey}  
\end{figure}

We think of the generated time series as the superposition of the motion protocol induced imprinted signal and the background turbulence. If we look at the frequency spectra as a variation of the inflow wind speed $u_{\inf}$ we find further indication of this circumstance. The turbulent power spectrum for five different wind speeds is shown in figure \ref{pic:spectrum_rey}. The frequency part of the imprinted signal up to $25~Hz$ is staying nearly constant for all wind speeds while the frequencies of the background turbulence are increasing and shifting with higher wind speeds. To determine the reproducibility of the generated turbulent wind speed time series this superposition has to be kept in mind. The analysis of the reproducibility of turbulent wind speed time series is done looking at the 50 wind speed time series acquired by the repetition of the presented motion protocol and measurement of the corresponding active grid wake. To examine the reproducible time scales and the influence of the superposition two filtering methods are used.\\

\subsubsection*{Filtering by moving average}
\begin{figure}
  \includegraphics[width=.4\textwidth]{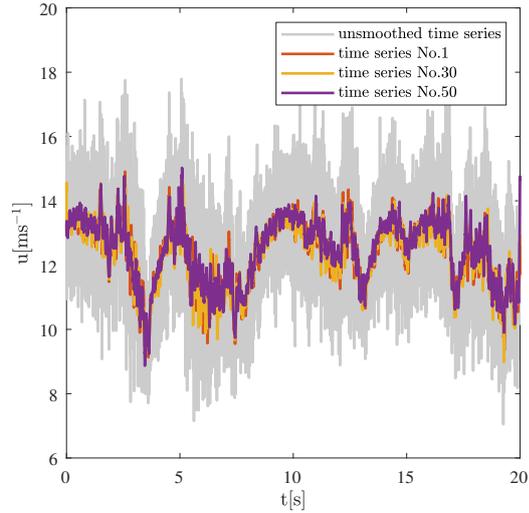}
\caption{Three exemplary wind speed time series generated by the motion protocol smoothed by a moving average filter for a better comparison.
An unfiltered wind speed time series is shown as reference in light gray.}
\label{pic:timeseries}      
\end{figure}

\begin{figure}
  \includegraphics[width=.4\textwidth]{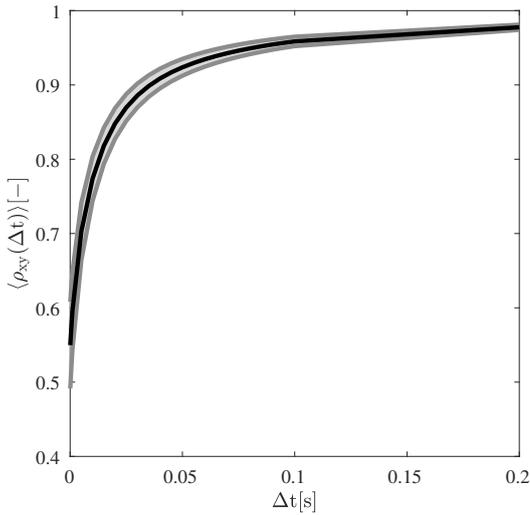}
\caption{Mean cross covariance coefficient of the 50 timeseries over the moving average span in seconds.}
\label{pic:xcov}      
\end{figure}

As a first glance at the reproducibility of the generated wind fields the data is compared plotting three of the acquired wind speed time series and using the reference filtering method of a moving average (fig. \ref{pic:timeseries}). The data is smoothed by a moving average filter using a subset of 800 samples. This corresponds to an averaging over $\frac{1}{25}~s$ and matches the highest frequency of the active grid motion protocol used.  
The three time series were chosen arbitrarily at the beginning, the middle and the end of the overall 50 acquired time series. In light gray, an unsmoothed time series is shown as a reference. Using a moving average filter the uncontrollable higher frequencies of the background turbulence are smoothed out and the underlying imprinted signal is becoming clearer. In this comparison a very good resemblance of the three examined time series can be observed for the chosen moving average span. Not only the trend but also many details for all the three time series are reproduced very well.\\
 An incremental increase and comparison of the resulting signals using the cross covariance function can give us a better impression about the reproducible scales that can be achieved. In figure \ref{pic:xcov} we see the mean of the maximum cross covariance function coefficient of all combinations of the 50 measurements over different moving average spans. The calculation of the mean cross covariance coefficient is decribed in appendix \ref{sec_methods_analysis}. 
As an envelope in light gray the standard deviation of the cross covariance coefficients for every span is added. Without averaging the coefficients we already get a value of over $0.55$. This indicates that the generated wind speed time series are containing recurring coherent structures. Also the standard deviation is already quite small. With enlarging moving average span $\Delta t$ the mean cross covariance coefficient $\langle \rho_{xy} \rangle$ is increasing rapidly till it is converging to a value significantly above $0.9$. Thus the coherence of the acquired wind speed time series can be viewed as very high as soon as the small fluctuations are filtered out via the moving average filtering. For the highest frequency of the motion protocol of $25~Hz$ the respective $\Delta t$ is $ 0.04~s$ and resulting in a cross covariance coefficient of $0.91$.\\
\subsubsection*{Filtering by empirical mode decomposition}
\begin{figure*} 
  \includegraphics[width=0.75\textwidth]{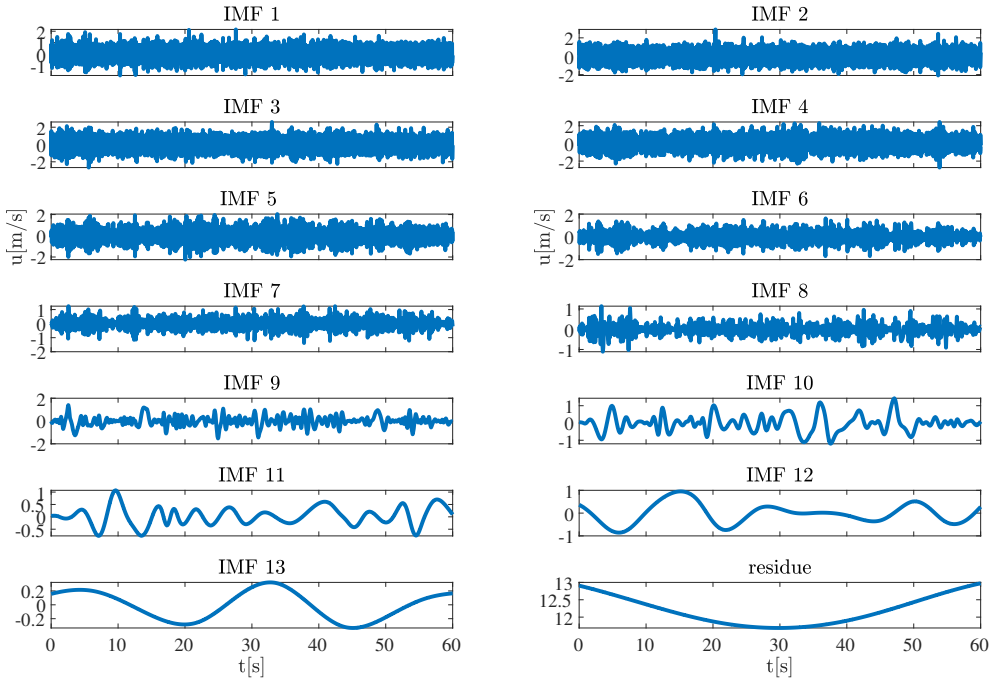}
\caption{First 13 intrinsic mode functions (IMFs) and residue of the empirical mode decomposition of one time series of the data set.}
\label{pic:IMFs}      
\end{figure*}

\begin{figure}
  \includegraphics[width=.4\textwidth]{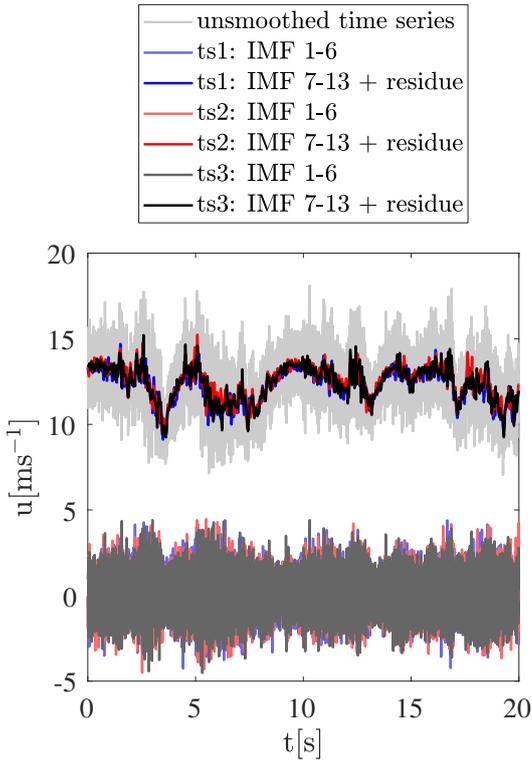}
\caption{Three exemplary wind speed time series generated by the LiDAR excitation protocol filtered by EMD. The imprinted signal (IMFs 7 to 13 + residue) is shown in darker colors fluctuating around $12 \frac{m}{s}$. The separated turbulent noise (IMFs 1 to 6 + residue) is shown in a lighter color fluctuating around $0 \frac{m}{s}$. An unfiltered wind speed time series is shown as reference in light gray.}
\label{pic:timeseries_emd}      
\end{figure}

\begin{figure}
  \includegraphics[width=.4\textwidth]{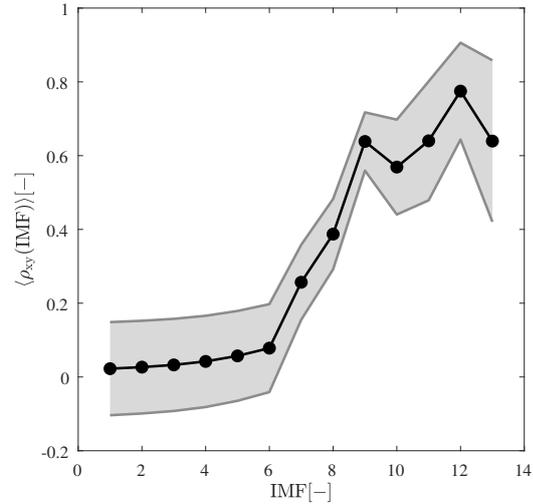}
\caption{Mean cross covariance coefficient of the 50 timeseries over the 13 IMFs. }
\label{pic:xcov_emd}     
\end{figure}

\begin{figure}
  \includegraphics[width=.4\textwidth]{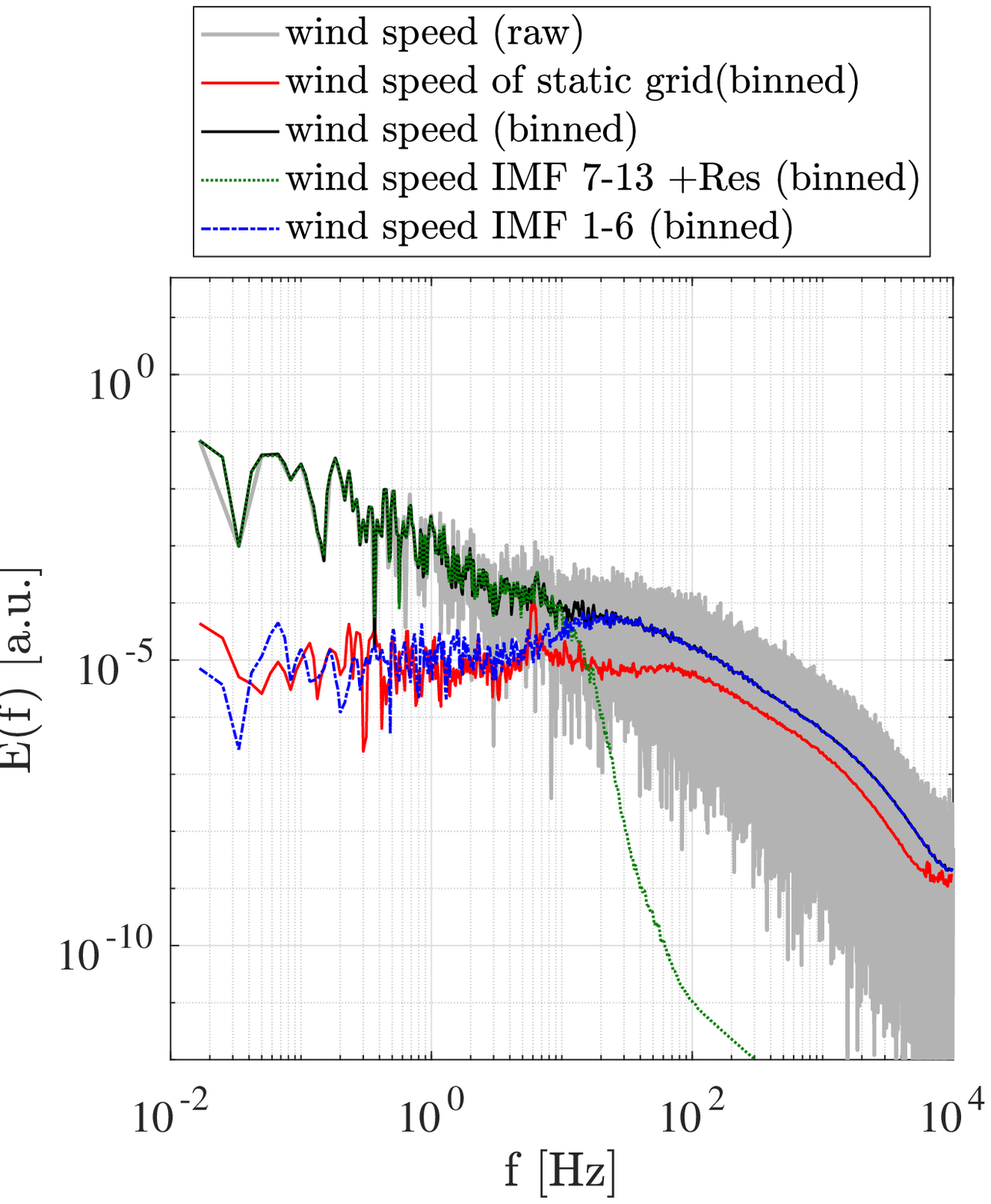}
\caption{Turbulent power spectra of the recombined time series using different IMFs of the EMD.}
\label{pic:spec_emd}    
\end{figure}

As indicated in appendix \ref{sec_methods_analysis}, a better way than a simple low pass filtering of the small scales of the signal would be the possibility to decompose the signal into the imprinted signal generated by the active grid motion protocol and the background turbulence induced by the desgin. This can be achieved by the empirical mode decomposition (EMD), as explained in appendix \ref{sec_methods_analysis}. In the EMD a time series is partitioned into several intrinsic mode functions (IMF) without leaving the time domain. Each IMF thereby correponds to a certain frequency domain of the original signal. In figure \ref{pic:IMFs} the first 13 intrinsic mode functions and the residue of the EMD of one time series of the data set are shown.\\
These IMFs can now be used to filter the measured time series in a more elaborate way. The recombination of the residue and several of the higher IMFs result in a filtered time series without the background turbulence. By combining the rest of the IMFs we get a time series of the background turbulence. This separation is shown in figure \ref{pic:timeseries_emd}.
For three exemplary time series the first six IMFs are combined to the background turbulence and modes 7 to 13 are combined with the residue to the filtered time series of the imprinted signal. These modes are chosen because this combination corresponds to a filtering at a frequency of around $13~Hz$ close to the moving average reference in figure \ref{pic:timeseries} and the highest frequency in the motion protocol. The filtered time series of the imprinted signal is varying around the mean wind speed of $12~\frac{m}{s}$ and the separated background turbulence is fluctuating around $0~ \frac{m}{s}$, see figure \ref{pic:timeseries_emd}. In a first brief comparison to the filtered time series in figure \ref{pic:timeseries} not much of a difference can be observed, as the same general trend can be observed. To check which IMFs form the unreproducible background turbulence and which the imprinted signal again the cross covariance function is used. For every IMF of the first 13 IMFs and all the 50 combinations the maximal cross covariance coefficient and the mean and standard deviation of the coefficient is calculated. The result is shown in figure \ref{pic:xcov_emd}. 
For the first six IMFs the cross covariance coefficient $\langle \rho_{xy} \rangle$ is around zero, which is expected for the higher frequent modes adding up to the uncoherent background turbulence. An increase of the cross covariance coefficient starts at mode 7 indicating the first occourring coherent structures. The highest values around a coefficient of $0.6$ are reached for the last four IMFs. By recombining the IMFs 7 to 13 we get a filtered time series containing all coherent structures imprinted by the active grid motion protocol.
\begin{figure}
  \includegraphics[width=.4\textwidth]{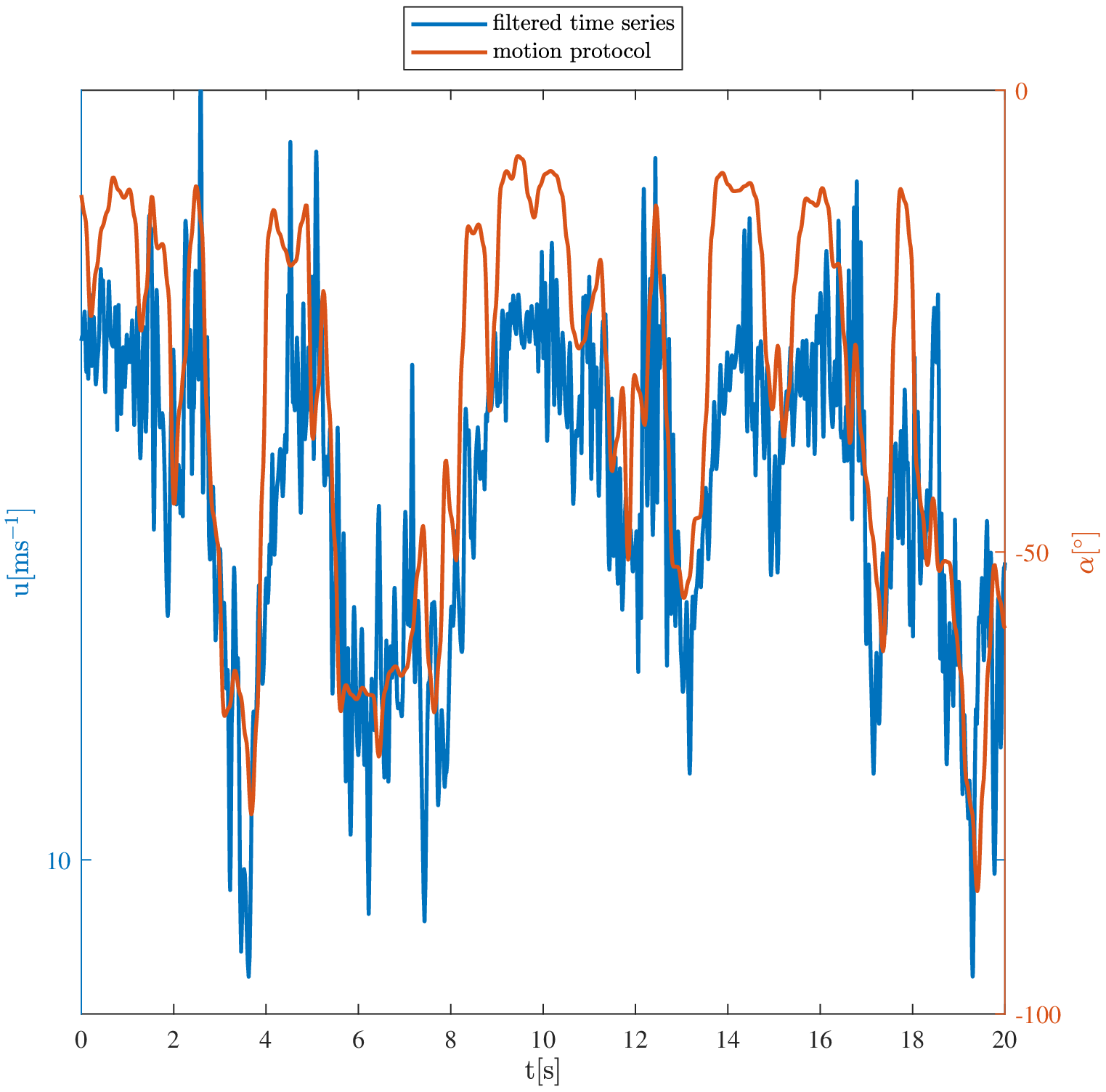}
\caption{Filtered wind speed time series via EMD (IMFs 7 to 13 + residue) and  flap angle fluctuation of the active grid motion protcol over time. A flap angle of 0/180 degrees correspons to minimal blockage and $\pm 90$ degrees to maximal blockage of the flow.}
\label{pic:protEMD}      
\end{figure}
This can be emphasized by a comparison to the absolute flap angle position of the active grid motion protcol, see figure \ref{pic:protEMD}. As described earlier an absolute flap angle $\alpha$ of $0/180^\circ$ corresponds to minimal blockage and $\pm 90^\circ$ to maximal blockage of the flow. An increase of blockage reduces therefore the wind speed. Thus the reproducible parts of our active grid wake are mainly depending on the change of the cross section blockage.\\ 
To determine the time scales the turbulent power spectrum can be used (figure \ref{pic:spec_emd}). 
Shown are the two resulting spectra when recombining the IMFs 1 to 6 and 7 to 13 plus the residue. Additionally in light grey an unfiltered spectrum, in black a binned untreated spectrum and in red a binned spectrum of the static active grid in fully open position are shown. From this figure we can deduce the following results. The reconstructed wind speed time series containing the IMFs 7 to 13 and the residue covers the whole range of frequencies up to $13~Hz$ and lower and then decreases rapidly. For this structures the flow can be reproduced accurately with a cross covariance coefficient of $0.9$ for the reconstructed time series using the modes 7 to 13.\\
The reconstructed time series containing the uncoherent high frequent IMFs is covering the rest of the untreated reference time series.  The spectrum containing the high frequent IMFs bare a good resemblance to the static reference, as it is also showing the $-\frac{5}{3}$ power decay law of Kolmogorov for the higher frequencies.\\

\subsection{Determining reproducibility based on statistic methods}\label{sec_statistic}
\begin{figure}
  \includegraphics[width=.4\textwidth]{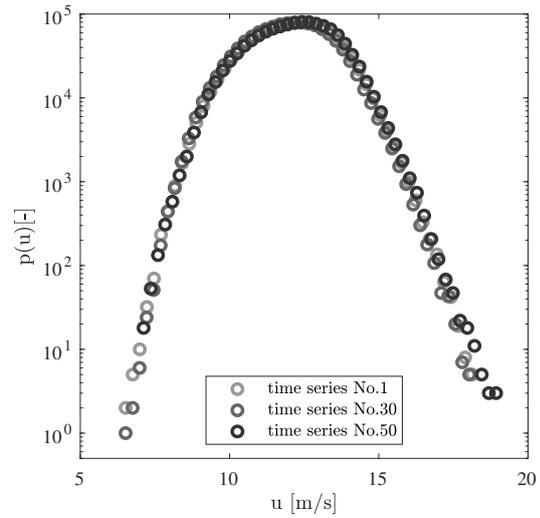}
\caption{PDFs of the wind speeds of the generated time series by the active grid.}
\label{pic:PDF_rep}     
\end{figure}

\begin{figure}
  \includegraphics[width=.4\textwidth]{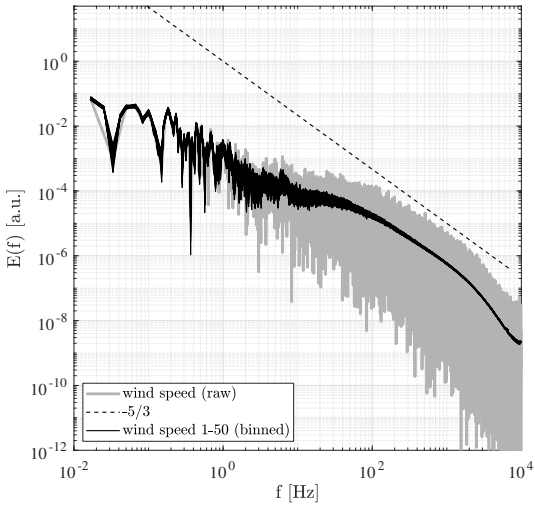}
\caption{Turbulent power spectra of all the 50 measurments smoothed by a windowing filtering.  An unfiltered turbulence spectrum is shown as reference in light gray.}
\label{pic:spectrum_all}    
\end{figure}

\begin{figure}
  \includegraphics[width=.4\textwidth]{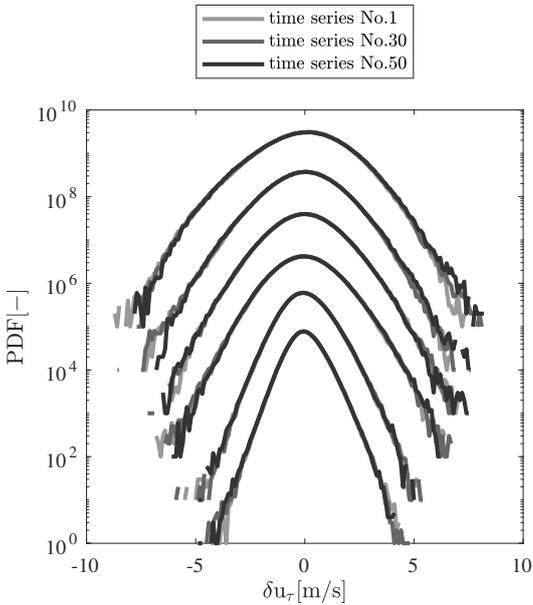}
\caption{PDFs of the wind speed increments for three exemplarily wind speed time series. PDFs are shifthed vertically for a better visibility. The increments are increasing from bottom to top ($\delta \tau = \frac{1}{2}~s,~ \frac{1}{20}~s,~\frac{1}{100}~s,~\frac{1}{200}~s,~\frac{1}{1000}~s,~ \frac{1}{2000}~s$)}
\label{pic:rep_inc}     
\end{figure}

\begin{figure}
  \includegraphics[width=.4\textwidth]{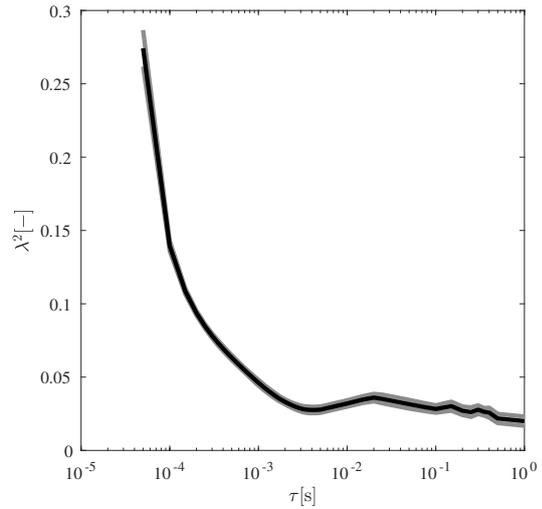}
\caption{Mean shape parameter $\lambda^{2}$ for all 50 measurements over the velocity increment spanwidth. Standard deviation of the mean in light grey as envelope.}
\label{pic:rep_lambda}    
\end{figure}

The last comparision is performed in this section, where statistic methods are used to determine the reproducibility further. The 50 measured time series are characterized based on one and two point statistics. The results are compared between the time series to determine the resemblance. First one point statistics are determined, followed by low and higher order two point statistics.\\
In table \ref{tbl:uTI} the average wind speed of the series of 50 measurements and the TI are shown with the corresponding standard deviation obtained from the 50 different measurements, and thus quantifying the accordance between the different time series. The standard deviation for both values is very low. Using this very simple statistic comparison the wind speed time series are matching pretty well to each other, which is not too surprising as the impact of the highly reproducible larger structures with larger amplitudes in the flow should be higher than the small structures with smaller amplitudes.\\
   \begin{table} [h]
   	     \caption{Mean wind speed and turbulence intensity of the 50 measurements with corresponding standard deviation.}
     \centering
     \begin{tabular}{l|c|c|c}
        $\langle u \rangle / \frac{m}{s}$ &$\sigma_{\langle u \rangle} / \frac{m}{s}$ & $TI / \%$ & $\sigma_{TI} /\%$ \\
       \hline
        $12.1$    & $0.1$&$11$ & $6*10^{-4}$         \\
     \end{tabular}

     \label{tbl:uTI}
   \end{table}
When comparing the probability density functions (PDF) of three exemplarily chosen wind speed time series a very good agreement can be seen in figure \ref{pic:PDF_rep}. The distribution of wind speeds in the time series varies scarcely and mainly for the highest and lowest wind speeds, which is caused by low count rates. 
Further analysis using two point statistics of the repeated wind speed time series is showing that also the higher time scales are statistically comparable like for static grids. In figure \ref{pic:spectrum_all} the frequency spectra of all 50 measurements are plotted on top of each other. 
In light grey an unfiltered reference of one spectra is shown. The repeated measurements are appearing nearly  throughout the whole frequency range as one single line. The low frequencies up to $2~Hz$ are induced mainly by the active grid motion and converge quickly for all 50 measurements. The main differences are in the range between $2~Hz$ and $50~Hz$, where we can observe a broadening of the line. On this scales the imprinted flow modulation is starting to decay into smaller structures which is a chaotic process. The structures are also still quite large compared to the measurement time of roughly $60~s$, so that the 50 spectra do not converge for that region. On the high frequency scales above $50~Hz$ the turbulence is self-sustaining and fully developed, since it follows the $-\frac{5}{3}$ law. Because of this the lines are converging again on this scales. Overall the resemblance of the spectra is quite high and we can speak of a good agreement.\\
Using the frequency spectra the statistical consistency of lowest order two point analysis could be shown. As we tried to generate atmospheric-like wind speed time series in our experiment and these are mostly intermittent, a further comparison of the two point statistics are performed using velocity increments. The definition of velocity increments can be found in appendix \ref{sec_methods_analysis}. The PDFs of the velocity increments complement the power spectra with respect to higher order two point statistics and are shown for three exemplary time series in \ref{pic:rep_inc}. 
Displayed are five different increments representing different time scales  $\delta \tau = \frac{1}{20}~s,~\frac{1}{100}~s,~\frac{1}{200}~s,~\frac{1}{1000}~s,~ \frac{1}{2000}~s$. The PDF are shifted vertically for better visibility, the smallest time scale at the bottom and the largest time scale at the top. The three depicted PDFs are in good agreement to each other and just minor differences are occurring for the rare extreme events, again due to the small number of counts. The shape parameter $\lambda^2$ is proportional to the normalized fourth central moment (kurtosis) and is used to indicate intermittency. The definition can be found in section \ref{sec_methods_analysis}. In figure \ref{pic:rep_lambda} the shape parameter is presented as the mean trend for all the 50 measurements (black) enveloped in light grey by the corresponding standard deviation. 
For increments of $2.5 \times 10^{-3}~s$ and smaller the shape parameter is significantly increasing and different from zero. This means the PDF is deviating from a Gaussian distribution and thereby is intermittent. Taking all 50 measurements into account the match is very good receiving a standard deviation with such low values.\\

\section{Conclusion}\label{sec_conclusion}
An experiment is presented to examine the reproducibility of a turbulent flow generated by an active grid. The process of downscaling a measured wind speed time series with atmospheric fluctuations for an experiment in the wind tunnel is shown. The possible frequency range which can be reproduced is hereby coupled to the frequency of the active grid motion protocol. By repeating this motion protocol the reproducibility of the generated wind speed time series in the wind tunnel can be examined.\\
The modulation of the flow in the frequency range between $0.1~Hz$ to $25~Hz$ is defining the reproducibility of coherent structures in the generated wind speed time series. This reproducibility of the exact occurrence of structures in the measured time series was analysed using two filtering techniques and the cross covariance function. For a lowpass filtering at $25~Hz$ using a moving average filter a maximal cross covariance coefficient of $0.91$ can be achieved when comparing all combinations of the 50 repeated measurements. When we use the EMD to seperate the imprinted signal containing the coherent and reproducible structures and the uncoherent and self sustaining background turbulence the reconstructed time series are seperated at a frequency of around $13~Hz$. Here for all combinations a maximal cross covariance coefficient of $0.9$ is reached. Both filtering methods show a high reproducibility of the generated wind speed time series in the frequency region of the flow modulation by the active grid motion protocol.\\
 For frequencies of $25~Hz$ and higher we observe the natural decay of turbulence following the $-\frac{5}{3}$ law. This natural decay is driven by the active grid motion on larger scales but mainly by the interaction of the design of the active grid and the inflowing wind. This can also be shown by a change of the main speed of the wind tunnel. This background turbulence is self sustaining and not reproducible in detail with respect to the time series, but reproduced well with respect to its statistical features.\\
An analysis using statistical methods for the comparison we find a good reproducibility by repeating the active grid motion for all statistical values as all results are converging very well. This behaviour is also not limited to a certain frequency range but for the full generated wind speed time series.\\
By this work we have shown that the active grid can be used to generate realistic inflow conditions in a wind tunnel like it is found in the atmospheric boundary layer. A scale can be selected down to which the turbulent structure can be reproduced in a controlled way. Most importantly these wind structures can be reproduced with high quality. This findings allows to make in a new quality experiments on the impact of turbulent wind structures on wind turbines. For example the reproducibility allows to repeat experiments with different control strategies and to optimise a control system.


\begin{acknowledgements} 
This work is part of the INNWIND.EU project supported by the Seventh Framework Programme (FP7) under Grant Agreement No. 308974. The authors are also thankful for the support by the Ministry for Science and
Culture of Lower Saxony through the funding initiative Nieders\"achsisches Vorab (project ventus
efficiens).\\
The measured LiDAR time series of wind data was provided by the WindScanner Research and Innovation team members Nikolas Angelou, Mikael Sjöholm and Torben Mikkelsen at the DTU Risoe campus as part of the Smart Blades project. The pre-processing of this data was done by Marijn van Dooren from ForWind.
Furthermore, we thank Nico Reinke and Gerrit Kampers for fruitful discussions.
\end{acknowledgements}

%
%



\appendix
\section{Analysis of the turbulent wind speed time series}\label{sec_methods_analysis} 
In this section the used methods to determine the reproducibility of the generated wind speed time series are described in detail.\\
The reproducibility of structures of the generated time series is examined using the cross covariance function. The resemblance of two time series can be presented as a single coefficient by calculating the cross covariance as a function of the lag and by determining the maximal value of the function ($\rho$). As a number of {\it n} measured time series are compared we get one coefficient for every comparison resulting in a matrix of coefficients $\rho_{ij}$ with $i,j:1..n$; {\it n} equals the number of measured time series (eq. \ref{eq:2}) 
\begin{equation}\label{eq:2}
  \ \rho =
  \begin{bmatrix}
   \rho_{1,1} & \dots & \rho_{1,n}\\
   \vdots & \ddots & \vdots \\
   \rho_{n,1} & \dots & \rho_{n,n}\\
  \end{bmatrix}.
\end{equation}
The diagonal coefficients in eq. \ref{eq:2} will become one as the autocovariance of one time series is calculated. The mean value and standard deviation of the matrix are now good measures to determine the resemblance for the whole data set of measurements.\\
The two filtering methods used are a simple moving average as reference and the empirical mode decomposition (EMD) introduced in the Huang Hilbert transformation \cite{Huang1998}. The moving average is a quick and common known filtering method to smooth a signal. But this method lacks a clear termination criteria for the here sought separation. While increasing the averaged subset of samples and comparing the result to similar time series the comparison only gets better as all smoothed time series are converging to the same mean value. Thus the moving average is used mainly as a reference for the second filtering method EMD.\\
 In EMD a time series get partitioned into several intrinsic mode functions (IMF) without leaving the time domain by an adaptive sifting process. The sifting process is an iterative process of enveloping the signal by determining the local minima and maxima. The mean of the lower and upper envelope forms the first IMF and gets subtracted from the signal. After that the next iteration begins. The sifting process repeats till just a residue with no local extrema is left or till a user-defined number of modes is reached. These IMFs can be used to bandpass filter the signal. A recombination of all IMFs results in the original time series but a recombination of the residue and only several of the last IMFs leaving out the first high frequent IMFs result in a smoothed filtered signal. The left out IMFs can be recombined to the high frequent noise. Overall, this procedure generates two time series representing the reproducible trend of the imprinted signal and the background turbulence. The finite number of IMFs can be recombined also just in a finite number of combinations.\\
Using the two filtering methods we get several matrices for $\rho$ for both ways of filtering: one for every subset of samples ($\langle \rho(\Delta t)\rangle$) used for the moving average filter and for all the combinations of 13 IMFs ($\langle \rho (\#n) \rangle$) of the EMD filtering. The correlation analysis by the just mentioned covariance shows how the direct time structures of the signal are reproduced.\\
%
 The wind speed time series are also analysed statistically to examine the turbulent features of the generated wind field. Common methods in wind energy science to investigate atmospheric wind data are one and two point statistics, which will be used here to that purpose.\\ 
One point statistics: A brief and fast way to compare the generated wind speed time series is the turbulence intensity (TI) defined as
\begin{equation}\label{eq:1}
TI = \frac{\sigma_{u}}{\langle u \rangle},
\end{equation}
where $\sigma_{u}$ denotes the standard deviation and $\langle u \rangle$ the mean of the analysed wind speed time series. The distribution of wind speeds in the time series is analysed using the probability distribution function (PDF) of the wind speeds.\\
Two point statistics: The simplest two point statistics method is the power spectrum. The frequency spectrum is used to determine the turbulent characteristics of the original and the generated wind speed time series. Further the slope of the spectrum is also a very good value to asses the resemblance of the statistical distribution of the frequencies. The frequency spectra are smoothed by a binning filter, which divides the spectra in equidistant windows on a logharithmic scale and averages the bins.\\
Furthermore the velocity increments of the time series are analysed. The velocity increment is defined as the subtraction of the wind speed at a certain time t and the wind speed with an added time shift $\tau$ (eq. \ref{eq:3}).
\begin{equation} \label{eq:3}
\delta u_\tau = u(t+\tau) - u(t)
\end{equation} 
The velocity increments are indicating how often gusts of different time scales are appearing in the investigated time series. For wind speed time series of the atmospheric boundary layer the PDFs of the velocity increments are the most general statistical aspect. The width of the PDF $\langle \delta u^{2}\rangle$ as a function of $\tau$ has the same information as the power spectrum. The deviation from the Gaussian distribution encompass the well known intermittency effect of small scale turbulence, which can also be expressed using the higher order structure function $\langle \delta u^{n} \rangle$. A normalized version of the kurtosis known as the shape parameter $\lambda^2$ (eq. 4) is used here to determine the intermittency of the investigated time series  \cite{chilla} 
\begin{equation}
\lambda^{2} = \ln \left( \frac{\langle \delta u^{4} \rangle}{3 \langle \delta u^{2}\rangle ^{2}} \right),
\end{equation}
knowing $\lambda^2$ and $\langle \delta u^{2}\rangle$ the function of the PDF is given \cite{castaing1990}.
Representative it should grasp the whole form of the PDF and thus in a compact way all higher moments, under the assumption that the higher moments are fitting as well when the fourth moment is matching.
%
%

\end{document}